\documentclass[conference]{IEEEtran}
\usepackage[cmex10]{amsmath}
\usepackage{amsfonts}
\usepackage[pdftex]{color,graphicx}
\usepackage{subfig}
\usepackage{latexsym}

\usepackage{xspace}
\newtheorem{thm}{Theorem}[section]  
\newtheorem{corol}[thm]{Corollary}
\newtheorem{defn}{Definition}[section]  
\newtheorem{eg}{Example}[section]

\usepackage[bookmarks=true,plainpages=false,pdfpagelabels,hypertexnames=false]{hyperref}
\hypersetup{
   colorlinks=true,
   pdftitle={},
   pdfauthor={Prabhakaran \& Prabhakaran},
   citecolor=black,
   linkcolor=blue
}

\newcommand{\namedref}[2]{\hyperref[#2]{#1~\ref*{#2}}}
\newcommand{\sectionref}[1]{\namedref{Section}{sec:#1}}

\newcommand{\definitionref}[1]{\namedref{Definition}{def:#1}}
\newcommand{\corollaryref}[1]{\namedref{Corollary}{cor:#1}}
\newcommand{\exampleref}[1]{\namedref{Example}{eg:#1}}

\newcommand{\figureref}[1]{\namedref{Figure}{fig:#1}}

\newcommand{\TODO}[1]{{#1}}

\newcommand{\Real}{\ensuremath{\mathbb{R}}\xspace}
\newcommand{\Rplus}{\ensuremath{{{\mathbb{R}}^+}}\xspace}

\newcommand{\CI}{\ensuremath{\text{\sf CI}}}
\newcommand{\RD}{\ensuremath{\text{\sf RD}}}
\newcommand{\CIO}{\ensuremath{\text{\sf CI-0}}}
\newcommand{\RDO}{\ensuremath{\text{\sf RD-0}}}
\newcommand{\sX}{\ensuremath{\mathcal X}}
\newcommand{\sY}{\ensuremath{\mathcal Y}}
\newcommand{\sQ}{\ensuremath{\mathcal Q}}
\newcommand{\sR}{\ensuremath{\mathcal R}}
\newcommand{\sRs}{\ensuremath{\mathcal R}_\star}
\newcommand{\bbZ}{\ensuremath{\mathbb Z}}
\newcommand{\sRsCI}{\ensuremath{{\mathcal R}_\CI}}
\newcommand{\sRsRD}{\ensuremath{{\mathcal R}_\RD}}
\newcommand{\sRCI}{\ensuremath{{\mathcal R}_{\star\CI}}}
\newcommand{\boundarysRsCI}{\ensuremath{\delta{\mathcal R}_{\CI}}}
\newcommand{\sRRD}{\ensuremath{{\mathcal R}_{\star\RD}}}
\newcommand{\CIWyner}{\ensuremath{C_\text{\sf Wyner}}}

\newcommand{\K}[3]{\ensuremath{K[{#1};{#2}|{#3}]}\xspace}
\newcommand{\Mfunc}{\ensuremath{\mathbb{M}}\xspace}
\newcommand{\M}[2]{\ensuremath{\Mfunc({#1};{#2})}\xspace}
\newcommand{\Kfunc}{\ensuremath{\mathbb{K}}\xspace}
\newcommand{\KK}[2]{\ensuremath{\Kfunc({#1};{#2})}\xspace}

\newcommand{\pt}{\ensuremath{\mathbf{a}}\xspace}

\newcommand{\pialice}{\ensuremath{\pi_{\mathrm{Alice}}}\xspace}
\newcommand{\pibob}{\ensuremath{\pi_{\mathrm{Bob}}}\xspace}
\newcommand{\Pialice}{\ensuremath{\Pi^{\mathrm{view}}_{\mathrm{Alice}}}\xspace}
\newcommand{\Pibob}{\ensuremath{\Pi^{\mathrm{view}}_{\mathrm{Bob}}}\xspace}
\newcommand{\Pialiceout}{\ensuremath{\Pi^{\mathrm{out}}_{\mathrm{Alice}}}\xspace}
\newcommand{\Pibobout}{\ensuremath{\Pi^{\mathrm{out}}_{\mathrm{Bob}}}\xspace}

\begin{document}

\title{Assisted Common Information with Applications to Secure Two-Party Computation}

\author{\IEEEauthorblockN{Vinod Prabhakaran and Manoj Prabhakaran}
\IEEEauthorblockA{University of Illinois, Urbana-Champaign\\
Urbana, IL 61801}}


\maketitle

\begin{abstract}
Secure multi-party computation is a central problem in modern cryptography. An important sub-class of this are problems of the following form: Alice and Bob desire to produce sample(s) of a pair of jointly distributed random variables. Each party must learn nothing more about the other party's output than what its own output reveals. To aid in this, they have available a {\em set up} --- correlated random variables whose distribution is different from the desired distribution --- as well as unlimited noiseless communication. In this paper we present an upperbound on how efficiently a given set up can be used to produce samples from a desired distribution.

The key tool we develop is a generalization of the concept of common information of two dependent random variables [G\'{a}cs-K\"{o}rner, 1973].  Our generalization --- a three-dimensional region --- remedies some of the limitations of the original definition which captured only a limited form of dependence. It also includes as a special case Wyner's common information [Wyner, 1975]. To derive the cryptographic bounds, we rely on a monotonicity property of this region: the region of the ``views'' of Alice and Bob engaged in any protocol can only monotonically expand and not shrink. Thus, by comparing the regions for the target random variables and the given random variables, we obtain our upperbound.

\end{abstract}

\section{Introduction}

Finding a meaningful definition for the ``common information'' of a pair of
dependent random variables $X$ and $Y$ has received much attention starting
from the
1970s~\cite{GacsKo73,Witsenhausen75,Wyner75,AhlswedeKo74,Yamamoto94}.  We
propose a new measure --- a three-dimensional region --- which brings out
a detailed picture of the extent of common information of a pair.  This
gives us an expressive means to compare different pairs with each other, based on
the shape and size of their respective regions.  We are motivated by
potential applications in cryptography, game theory, and distributed
control, besides information theory, where the role of dependent random
variables and common randomness is well-recognized.

\begin{figure}[tb]
\centering
\scalebox{0.23}{\includegraphics{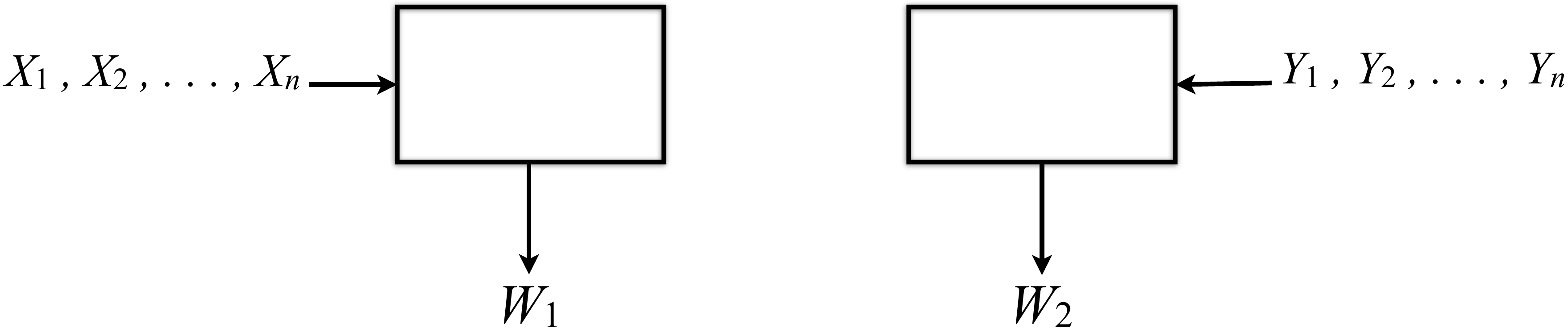}}\\(a)\\\vspace{0.1cm}
\scalebox{0.23}{\includegraphics{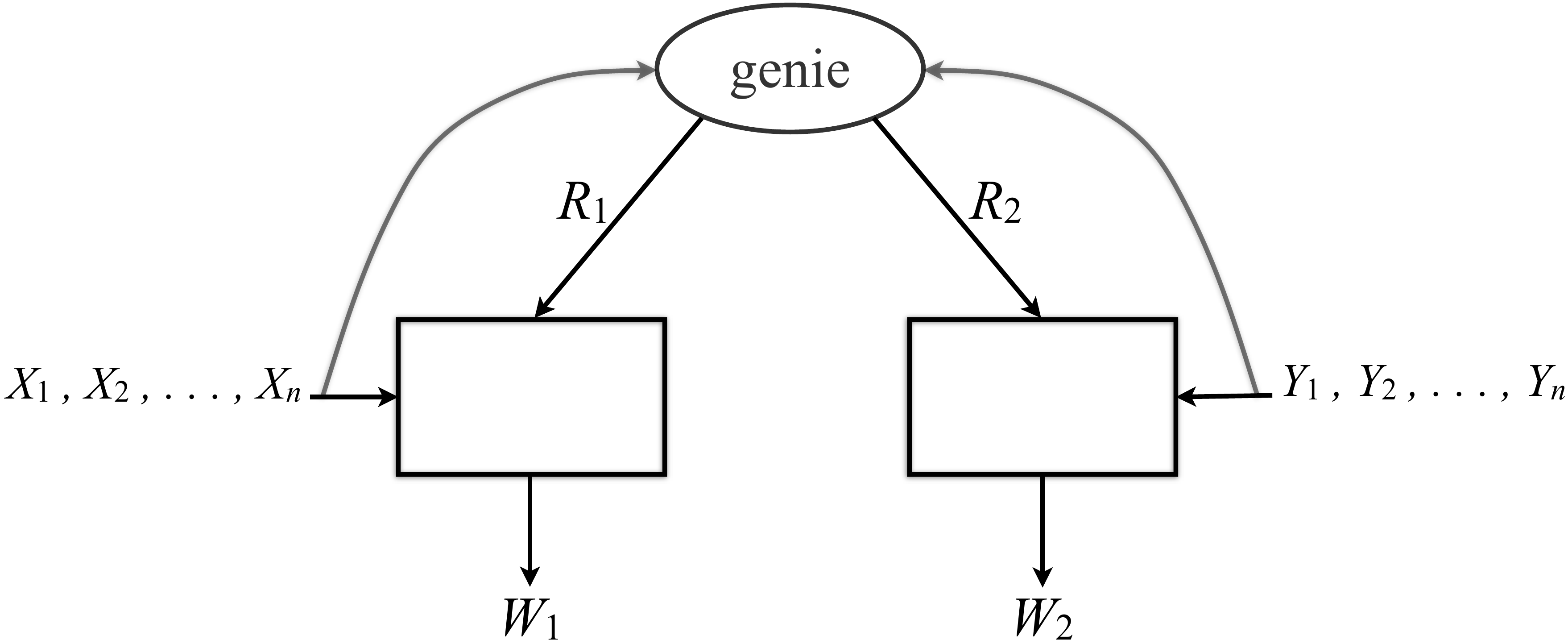}}\\(b)
\caption{Setup for (a) G\'{a}cs-K\"{o}rner common information, and
(b) assisted common information.}
\label{fig:setup}
\end{figure}

Suppose $X=(X',Q)$ and $Y=(Y',Q)$ where $X',Y',Q$ are independent. Then a
natural measure of ``common information'' of $X$ and $Y$ is $H(Q)$. Both an
observer of $X$ and an observer of $Y$ may independently produce the common
part $Q$; and conditioned on $Q$, there is no ``residual dependency,''
i.e., $I(X;Y|Q)=0$. The definition of G\'{a}cs and
K\"{o}rner~\cite{GacsKo73} generalizes this to arbitrary $X,Y$
(Fig.~\ref{fig:setup}(a)): the two observers now see $X^n=(X_1,\ldots,X_n)$
and $Y^n=(Y_1,\ldots,Y_n)$, resp., where $(X_i,Y_i)$ pairs are independent
drawings of $(X,Y)$. They are required to produce random variables
$W_1=f_1(X^n)$ and $W_2=f_2(Y^n)$, resp., which agree (with high
probability). The largest entropy rate (i.e., entropy normalized by $n$) of
such a ``common'' random variable was proposed as the common information of
$X$ and $Y$. However, in the same paper~\cite{GacsKo73}, G\'{a}cs and
K\"{o}rner showed (a result later strengthened by
Witsenhausen~\cite{Witsenhausen75}) that this rate is still just the
largest $H(Q)$ for $Q$ such that $X$ and $Y$ can be written as $(X',Q)$ and
$(Y',Q)$ respectively.%
\footnote{Hence, after removing the maximal such $Q$, the contribution to
the common information from $X'$ and $Y'$ is zero, even if they are highly
correlated. Other approaches which do not necessarily suffer from this
drawback have been suggested, 
notably~\cite{Wyner75,AhlswedeKo74,Yamamoto94}.}
In other words, this
definition captures only an explicit form of common information in (a
single instance of) $(X,Y)$.

One limitation of the common information defined by G\'{a}cs and K\"{o}rner
is that it ignores information which is {\em almost} common. Our approach
could be viewed as a strict generalization of theirs which uncovers extra
layers of ``almost common information.'' Technically, we introduce an
omniscient genie who has access to both the observations $X$ and $Y$ and can
send separate messages to the two observers over rate-limited noiseless
links.  See Fig.~\ref{fig:setup}(b). The objective is for the observers to
agree on a ``common'' random variable as before, but now with the genie's
assistance.  This leads to a trade-off region trading-off the rates of the
noiseless links and the resulting common information\footnote{We use the
term common information primarily to maintain continuity
with~\cite{GacsKo73}.} (or the resulting residual dependency). We
characterize these trade-off regions and show that, in general, they exhibit
non-trivial behavior, but reduce to the trivial behaviour discussed above
when the rates of the noiseless links are zero.

Our new measure has an immediate application to cryptography
(\sectionref{crypto}). Distributed
random variables with non-trivial correlations form an important resource in
the cryptographic task of secure multi-party computation. A fundamental
problem here is for two parties to ``securely generate'' a certain pair of
random variables, given another pair of random variables, by means of a
protocol. We show that the region of residual dependency of the views of two
parties engaged in such a protocol can only monotonically expand and not
shrink. Thus, by comparing the regions for the target random variables and
the given random variables, we obtain 
improved upperbounds on the efficiency with which one pair can be used to
securely generate another pair.

\section{Assisted Common Information Region}

\subsection{Characterization}

We say that a rate pair $(R_1,R_2)$ {\em enables} a common information
rate $R_\CI$ if for every $\epsilon>0$, there is a large enough integer $n$
and (deterministic) functions $f_k:\sX^n\times\sY^n \rightarrow \{1,\ldots,
2^{n(R_k+\epsilon)}\}$, $(k=1,2)$, $g_1:\sX^n\times\{1,\ldots,
2^{n(R_1+\epsilon)}\} \rightarrow \bbZ$, and $g_2:\sY^n\times\{1,\ldots,
2^{n(R_2+\epsilon)}\} \rightarrow \bbZ$ (where $\bbZ$ is the set of integers)
such that
\begin{align}
&\Pr\left( g_1(X^n,f_1(X^n,Y^n)) \neq g_2(Y^n,f_2(X^n,Y^n)) \right) \leq
\epsilon,\label{eq:prob-of-error}\\
&\frac{1}{n}I(X^n,Y^n;g_1(X^n,f_1(X^n,Y^n))) \geq R_\CI - \epsilon.
\label{eq:CIrate}
\end{align}
We denote the closure of the set of all rate pairs which enable 
a common information rate $R_\CI$ by $\sRsCI(R_\CI)$. We call this the
{\em rate-region} for enabling a common information rate of $R_\CI$.
Note that the largest value of $R_\CI$ we need consider is $H(X,Y)$. For
larger values of $R_\CI$, $\sRsCI(R_\CI)$ is clearly empty.

Similarly, we define the {\em rate-region} $\sRsRD(R_\RD)$ for enabling a
residual dependency rate of $R_\RD$ as the closure of the set of all rate
pairs which enable a residual dependency rate $R_\RD$, where the definition
of what it means for a rate pair to {\em enable} a residual dependency rate
$R_\RD$ is exactly as above except \eqref{eq:CIrate} is replaced by
\begin{align*}
\frac{1}{n} I(X^n;Y^n|g_1(X^n,f_1(X^n,Y^n))) \leq R_\RD + \epsilon.
\end{align*}
We also define the following ``single-letter'' regions
{\small 
\begin{align}
&\sRCI(R_\CI) = \left\{ (I(Y;Q|X),I(X;Q|Y)) : I(X,Y;Q) \geq
R_\CI\right\},\label{eq:CIregion}\\
&\sRRD(R_\RD) = \left\{ (I(Y;Q|X),I(X;Q|Y)) : I(X;Y|Q) \leq
R_\RD\right\}.\label{eq:RDregion}
\end{align}}
Here $Q$ is any random variable dependent on $(X,Y)$.

The main result of this section is a characterization of the rate-regions
defined above
(proof is sketched in section~\ref{subsec:proof}):
\begin{thm} \label{thm:main}
\begin{align}
\sRsCI&=\sRCI,\\
\sRsRD&=\sRRD.
\end{align}
Further, the cardinality of the alphabet $\sQ$ of $Q$ in
\eqref{eq:CIregion}-\eqref{eq:RDregion} can be restricted
to $|\sX||\sY|+2$.
\end{thm}
\TODO{}

\subsection{Behavior at $R_1=R_2=0$ and Connection to
G\'{a}cs-K\"{o}rner~\cite{GacsKo73}}

As discussed in the introduction, G\'{a}cs-K\"{o}rner showed that when
there is no genie, the common information rate is zero unless $X=(X',Q)$,
$Y=(Y',Q)$, and $H(Q)>0$. Since the absence of links from the genie is a
more restrictive condition than zero-rate links from the genie, we can ask
whether introducing an omniscient genie, but with {\em zero-rate} links to
the observers, changes the conclusion of G\'{a}cs-K\"{o}rner.  The corollary
below answers this question in the negative.  Also note that the result of
G\'{a}cs-K\"{o}rner can be obtained as a simple consequence of this corollary.
\begin{align*}
\text{Let}\qquad
R_\CIO&=\sup\; \{R_\CI: (0,0)\in\sRsCI(R_\CI)\}, \text{ and}\\
R_\RDO&=\inf\; \{R_\RD: (0,0)\in\sRsRD(R_\RD)\}.
\end{align*}
\begin{corol} \label{cor:GacsKo}
$R_\CIO > 0$ (or, $R_\RDO< I(X;Y)$) only if there are $X',Y',Q'$
such that $X=(X',Q')$, $Y=(Y',Q')$, $R_\CIO=H(Q')$, and
$R_\RDO=I(X;Y|Q')$.
\end{corol}
\begin{IEEEproof}[Proof sketch.] We first observe that the only $Q$'s
allowed in \eqref{eq:CIregion} and \eqref{eq:RDregion} if the rate pair
$(0,0)$ is a member are such that $I(Q;Y|X)=I(Q;X|Y)=0$. Thus, the joint
p.m.f. of $X,Y,Q$ has the form\[ p(x,y,q)=p(x,y)p(q|x)=p(x,y)p(q|y).\]
Hence, for all $(x,y)$ such that $p(x,y)>0$, we must have
$p(q|x)=p(q|y)$, $\forall q$. This implies that, if we consider the
bipartite graph with vertices in $\sX\cup\sY$ and an edge between $x\in\sX$
and $y\in\sY$ if and only if $p(x,y)>0$, for all vertices in the same
connected component, $p(q|\text{vertex})$ is the same. Using this, and
defining $Q'$ to be the connected component to which $X$ (or, equivalently
$Y$) belongs, we can show that
\begin{align*}
I(X,Y;Q) &= I(Q';Q) \leq H(Q'),\\
I(X;Y|Q) &= H(Q'|Q) + I(X;Y|Q') \geq I(X;Y|Q').
\end{align*}
If there is only one connected component, this implies that $R_\CIO=0$ and
$R_\RDO = I(X;Y)$. Hence, if $R_\CIO>0$ (or, $R_\RDO < I(X;Y)$), more than
one connected component must exist; moreover $R_\CIO=H(Q')$ and 
$R_\RDO=I(X;Y|Q')$.
\end{IEEEproof}

Thus, at zero rates, common information exhibits trivial behavior. However, for
positive rates, the behavior is, in general, non-trivial. Presently, we will
demonstrate this through a few examples. But before that, we will show that
Wyner's common information can also be obtained as a special case of our
characterization.

\subsection{Connection to Wyner's Common Information~\cite{Wyner75}}

Wyner offered an alternative definition for common
information in~\cite{Wyner75}. Briefly, Wyner's common information is the
``minimum binary rate of the common input to two independent processors that
generate an approximation to $X,Y$.'' From~\cite{Wyner75}, Wyner's common
information is
\begin{align*}
\CIWyner =\inf I(X,Y;U),
\end{align*}
where the infimum is taken over $U$ such that $X - U - Y$ is a Markov
chain. It is easy to show that $\CIWyner\geq I(X;Y)$. Wyner's common
information can be obtained as a special case of our characterization:
(proof omitted due to space constraints)
\begin{corol}
\begin{align*}
\CIWyner - I(X;Y) = \min_{(R_1,R_2)\in\sRsRD(0)} R_1 + R_2.
\end{align*}
\end{corol}

\subsection{Non-Trivial Behavior at Non-Zero Rates}
\begin{figure}[tb]
\centering
\scalebox{0.32}{\includegraphics{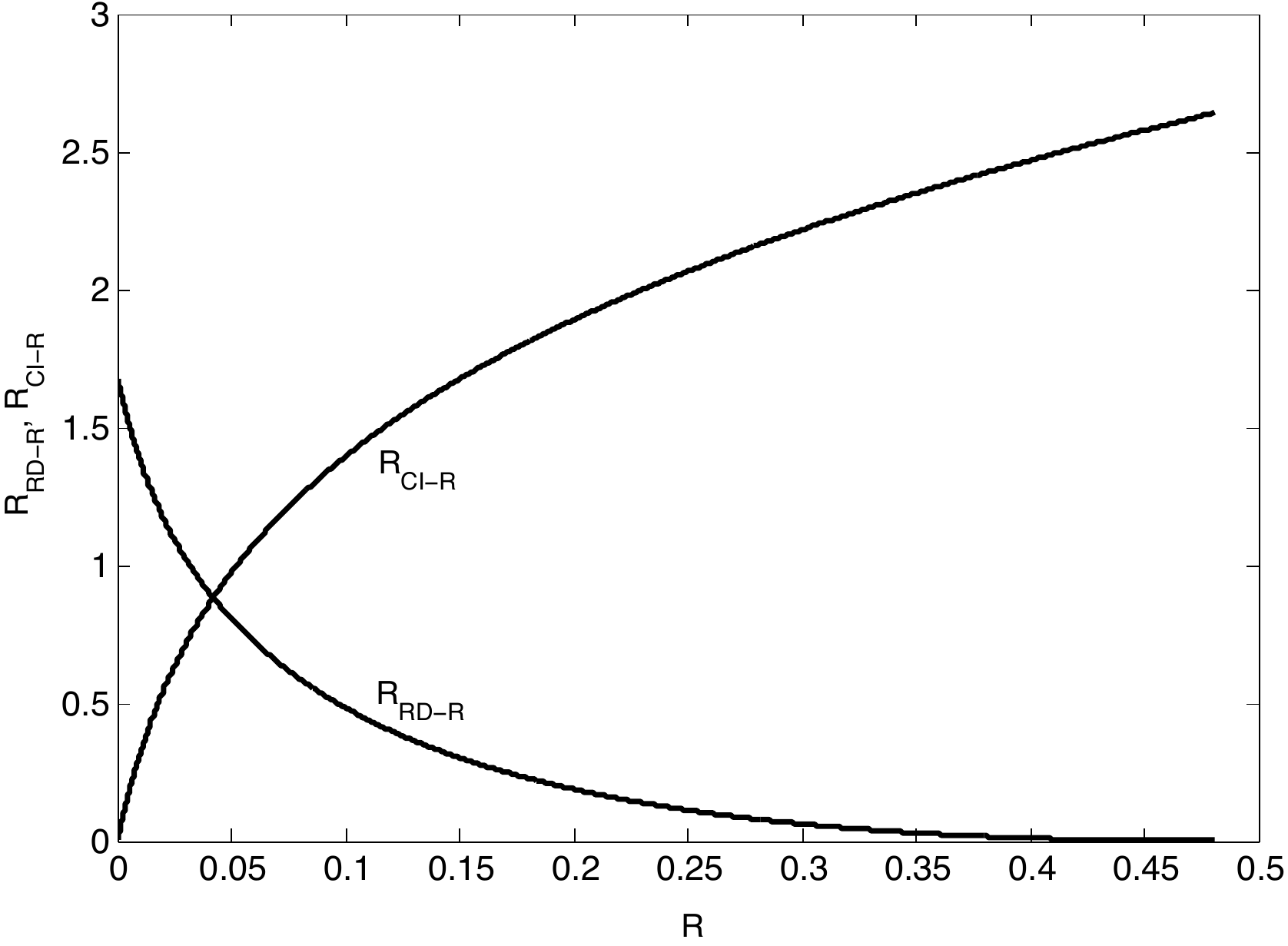}}
\caption{\small An achievable trade-off between $R_1=R_2=R$ and $R_\CI$ (also
$R_\RD$) for jointly Gaussian $X,Y$ of unit variance and correlation
$\rho=0.95$. The trade-off is obtained by choosing $Q$ in
\eqref{eq:CIregion} and \eqref{eq:RDregion} to be the optimal jointly
Gaussian choice. The optimal $R_\CI$ is at least as much as shown 
and the optimal $R_\RD$ is at most what is shown. Note that $R_\CI$
is strictly positive for all $R>0$.}
\label{fig:Gaussian}
\end{figure}
\begin{figure}[tb]
\centering
\scalebox{0.3}{\includegraphics{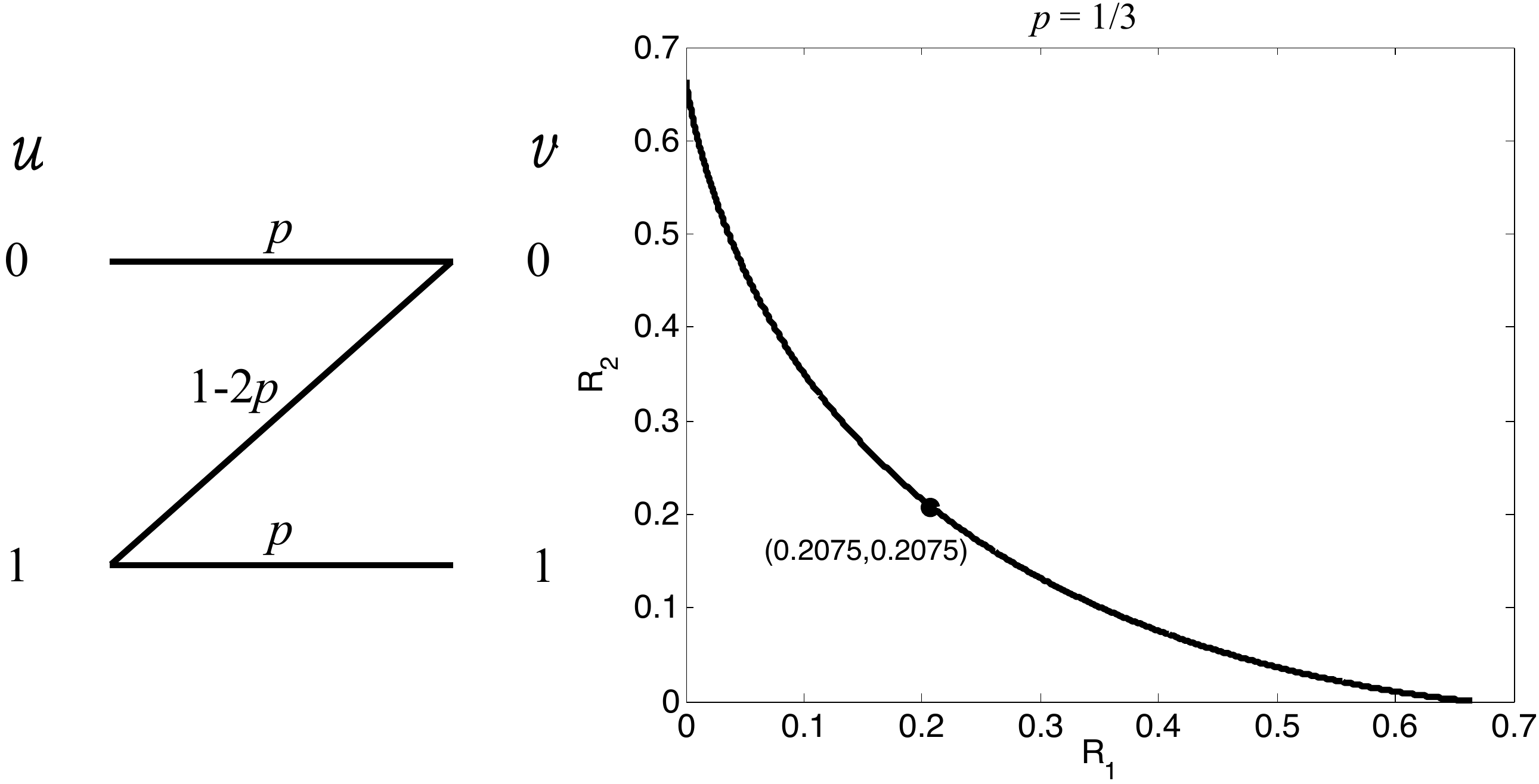}}
\caption{\small 
$U,V$ are binary random variables with joint p.m.f.
$p(0,0)=p(1,1)=p$, $p(1,0)=1-2p$, and $p(0,1)=0$.
Boundary of $\sRsRD(0)$ for $p=1/3$ is shown. The marked point is the
minimum sum-rate point.}
\label{fig:zsource}
\end{figure}
\begin{figure}[tb]
\centering
\scalebox{0.3}{\includegraphics{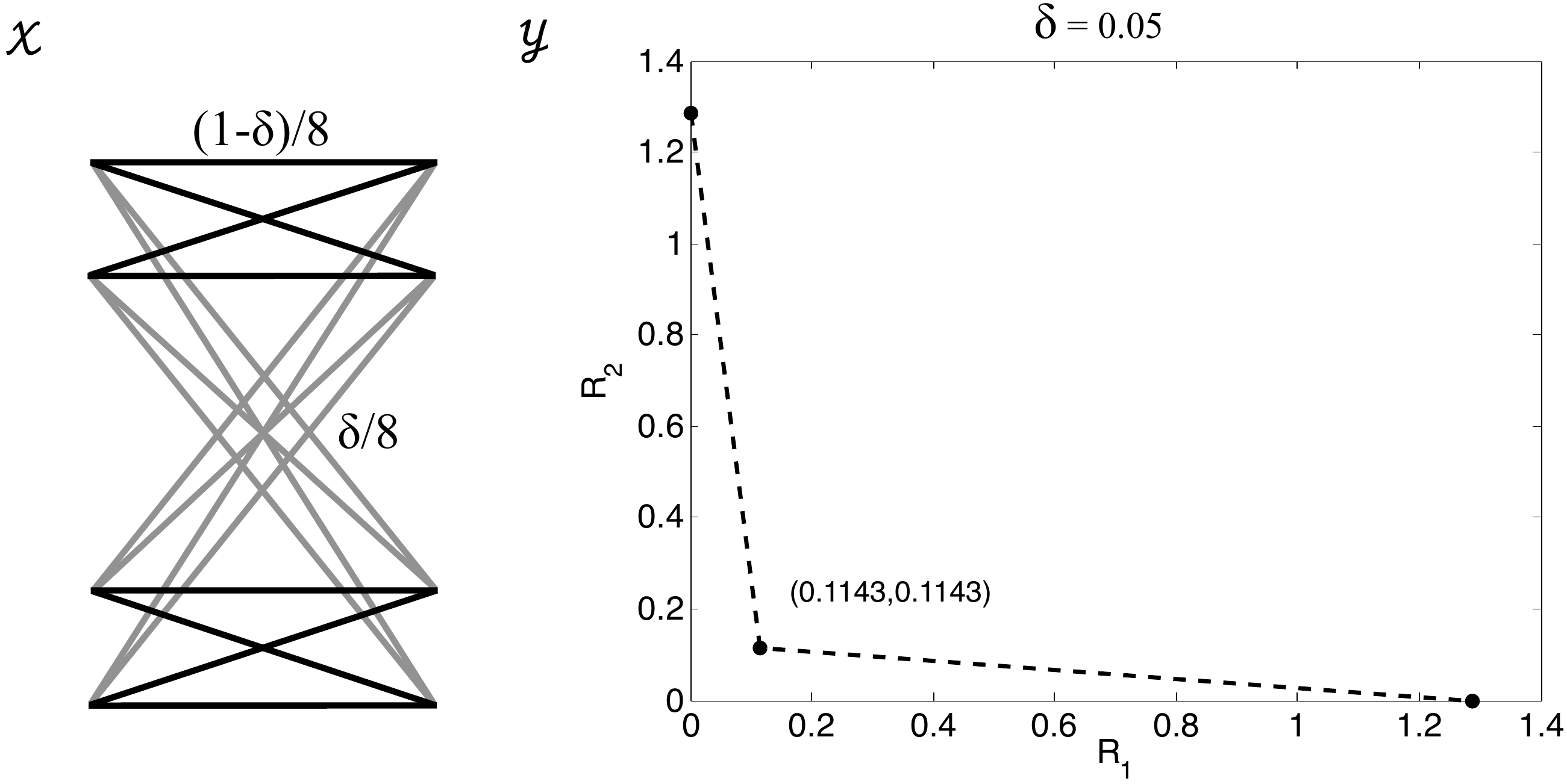}}
\caption{\small \TODO{$X,Y$ are dependent random variables whose joint
p.m.f is shown. The solid lines each carry a probability mass of
$\frac{1-\delta}{8}$ and the lighter ones $\frac{\delta}{8}$. In the plot,
all points on the dotted lines belong to $\sRsRD(0)$.}}
\label{fig:connected}
\end{figure}

\begin{eg}{\em Jointly Gaussian random variables.} \label{eg:Gaussian} We
consider jointly Gaussian\footnote{While the discussion has been for
discrete random variables, it extends directly to continuous random
variables.} $X,Y$ each of unit variance and with correlation coefficient
$\rho$.  Let the rates of the links from the genie to the two observers be
the same, $R_1=R_2=R$.  \figureref{Gaussian} plots an achievable $R_\CI$
and $R_\RD$ by choosing $Q$ in \eqref{eq:CIregion} and \eqref{eq:RDregion}
to be the optimal jointly Gaussian choice (jointly Gaussian with $X,Y$);
i.e, the optimal $R_\CI$ is at least as much as shown and the optimal
$R_\RD$ is at most what is shown. Note that $R_\CI=0$ when $R=0$ consistent
with Corollary~\ref{cor:GacsKo}, but $R_\CI$ is strictly positive for all
$R>0$. \end{eg}

\begin{eg}{\em A binary example.} \label{eg:zsource} \figureref{zsource}
shows the joint p.m.f. of a pair of dependent binary random variables
$U,V$.
The boundary of the rate region $\sRsRD(0)$ is plotted in
\figureref{zsource}. This is the optimal trade-off of rates at which the
genie can communicate with the observers so that they may produce a common
random variable which can render their observations practically
conditionally independent. \end{eg}

\TODO{\begin{eg} \label{eg:connected} \figureref{connected} shows the joint
p.m.f. of a pair of dependent random variables $X,Y$. When $\delta=0$, they
have the simple dependency structure of $X=(X',Q), Y=(Y',Q)$ where
$X',Y',Q$ are independent. This is the trivial case in the introduction,
and the observers can each produce, without any assistance from the genie,
$Q$ which renders their observations conditionally independent. Thus,
$\sRsRD(0)$ is the entire positive quadrant. For small values of $\delta$
we intuitively expect the random variables to be ``close'' to this case. A
measure such as the common information of G\'{a}cs and K\"{o}rner fails to
bring this out (common information is discontinuous in $\delta$ jumping
from $H(Q)=1$ at $\delta=0$ to 0 for $\delta>0$). However, the intuition is
borne out by our trade-off regions. For instance, for $\delta=0.05$,
\figureref{connected} shows that $\sRsRD(0)$ is nearly all of the positive
quadrant.
\end{eg}}

In \sectionref{crypto}, we will use the characterization developed in this
section to compare the pairs of random variables in the last two examples
in a cryptographic context. See \exampleref{crypto-example}.

\subsection{Relationship between $\sRsCI$ and $\sRsRD$}
The residual dependency rate-region can be written in terms of the common
information rate-region: (proof is omitted due to space constraints)
\begin{corol}
{\small \begin{align*}
\sRsRD(R_\RD)=\{ &(R_1,R_2): \exists (r_1,r_2)\in
    \boundarysRsCI(r_\CI) \text{ s.t. } r_\CI \geq\\&I(X;Y)-R_\RD+r_1+r_2,
    R_1\geq r_1, \text{ and } R_2\geq r_2\},\\
\intertext{where}
\boundarysRsCI(R_\CI)=\{&(R_1,R_2)\in\sRsCI(R_\CI): \not{\exists}
(r_1,r_2)\in\sRsCI(R_\CI)\\ &\text{ s.t. } r_1\leq R_1, r_2\leq R_2, 
\text{ and } (r_1,r_2)\neq (R_1,R_2)\}.
\end{align*}}
\end{corol}

\subsection{Sketch of Proof of Theorem~\ref{thm:main}} \label{subsec:proof}
Proof of achievability ($\sRs \supseteq \sR$), which is based on Wyner-Ziv's
source coding with side-information~\cite{WynerZi73}, is omitted in the
interest of space. The cardinality bound can be shown using Carath\'{e}odory's
theorem.

To prove the converse, let $\epsilon>0$ and $n,f_1,f_2,g_1,g_2$ be such that
\eqref{eq:prob-of-error} and \eqref{eq:CIrate} hold. Let
$C_k=f_k(X^n,Y^n)$, for $k=1,2$, and $W_1=g_1(X^n,C_1)$ and
$W_2=g_2(Y^n,C_2)$. Then,
{\small
\begin{align*}
R_1 + \epsilon &\geq \frac{1}{n} H(C_1) \geq \frac{1}{n} H(C_1|X^n)
 \geq \frac{1}{n} H(W_1|X^n)\\
 &\geq \frac{1}{n} I(Y^n;W_1|X^n)\\
 &\stackrel{(a)}{=} 
   \frac{1}{n} \sum_{i=1}^n H(Y_i|X_i) - H(Y_i|Y^{i-1},X^n,W_1)\\
 &\geq \frac{1}{n} \sum_{i=1}^n H(Y_i|X_i) -
H(Y_i|X_i,W_1,Y^{i-1},X^{i-1})\\
 &= \sum_{i=1}^n \frac{1}{n} I(Y_i;Q_i|X_i),\;Q_i:=(W_1,Y^{i-1},X^{i-1})\\
 &\stackrel{(b)}{=} I(Y_J;Q_J|X_J,J) \stackrel{(c)}{=} I(Y_J;Q|X_J),\;Q:=(Q_J,J),
\end{align*}}
where (a) follows from the independence of $(X_i,Y_i)$ pairs across $i$. In
(b), we define $J$ to be a random variable uniformly distributed
over $\{1,\ldots,n\}$ and independent of $(X^n,Y^n)$. And (c) follows from
the independence of $J$ and $(X^n,Y^n)$. Similarly,
{\small
\begin{align*}
R_2 + \epsilon &\geq \frac{1}{n} H(C_2|Y^n) \geq \frac{1}{n} H(W_2|Y^n) \\
 &\geq \frac{1}{n} H(W_1|X^n) - \frac{1}{n} H(W_2|W_1)\\
 &\stackrel{(a)}{\geq} H(W_1|X^n) - \kappa\epsilon\\
 &\geq\frac{1}{n} I(X^n;W_1|Y^n) - \kappa\epsilon\\
 &\stackrel{(b)}{\geq} I(X_J;Q|Y_J) - \kappa\epsilon,
\end{align*}}
where (a) (with $\kappa:=1+\log|\sX||\sY|$) follows from Fano's inequality and
the fact that the range of $g_1$ can be restricted without loss of
generality to a set of cardinality $|\sX|^n|\sY|^n$. And (b) can be shown
along the same lines as the chain of inequalities which gave a lower bound
for $R_1$ above. Moreover,
{\small
\begin{align*}
\frac{1}{n} I(X^n,Y^n;W_1) &= 
\frac{1}{n} \sum_{i=1}^n H(X_i,Y_i) -  H(X_i,Y_i|W_1,X^{i-1},Y^{i-1})\\
  &= \frac{1}{n} \sum_{i=1}^n I(X_i,Y_i;Q_i)\\
  &= I(X_J,Y_J;Q). 
\end{align*}}
Since $X_J,Y_J$ has the same joint distribution as $X,Y$, the converse
($\sRsCI\subseteq\sRCI$) for common information follows. Similarly, the converse
($\sRsRD\subseteq\sRRD$) for residual dependency can be shown using
{\small
\begin{align*}
\frac{1}{n} I(X^n;Y^n|W_1) &= \frac{1}{n} \sum_{i=1}^n I(X_i;Y^n|W_1,X^{i-1})\\
 &\geq \frac{1}{n} \sum_{i=1}^n I(X_i;Y_i|W_1,X^{i-1},Y^{i-1})\\
 &= I(X_J;Y_J|Q).
\end{align*}}


\section{Cryptographic Application}
\label{sec:crypto}

\subsection{Background}
\label{sec:cryptobg}
Secure multi-party computation is a central problem in modern cryptography.
Roughly, the goal of secure multi-party computation is to carry out
computations on inputs distributed among two (or more) parties, so as to provide
each of them with no more information than what their respective inputs and outputs
reveal to them. Our focus in this section is on an important sub-class of
such problems --- which we shall call {\em secure 2-party sampling} --- in which
the computation has no inputs, but the outputs to the parties are required to be
from a given joint distribution (and each party should not learn anything
more than its part of the output). Also we shall restrict ourselves to the
case of honest-but-curious adversaries.  It is well-known (see for instance
\cite{Wullschleger08thesis} and references therein) that very few distributions can be sampled from in this way,
unless the computation is aided by a {\em set up} --- some correlated random
variables that are given to the parties at the beginning of the protocol.
The set up itself will be from some distribution $(X,Y)$ (Alice gets $X$ and
Bob gets $Y$) which is different from the desired distribution $(U,V)$
(Alice getting $U$ and Bob getting $V$).
The fundamental question then is, which set ups $(X,Y)$ can be used to securely
sample which distributions $(U,V)$, and {\em how efficiently}.

While the feasibility question can be answered using combinatorial analysis
(as, for instance, was done in \cite{Kilian00}), information theoretic tools
have been put to good use to show bounds on efficiency of protocols (e.g.
\cite{Beaver96,DodisMi99,WinterNaIm03,ImaiMuNaWi04,WolfWu08,ImaiMoNa06,CsiszarAh07,WinklerWu09}).
Our work continues on this vein of using
information theory to formulate and answer efficiency questions in
cryptography. Specifically, {the quantities explored in the previous section}
lead to effective tools in providing new and improved upper-bounds on the rate
at which samples from a distribution $(U,V)$ can be securely generated, per
sample drawn from a set up distribution $(X,Y)$. Below we sketch the outline
of this application, which is further developed in \cite{MajiPrPrRo10}.

\paragraph{Secure Protocols}
A two-party protocol $\Pi$ is specified by a pair of (possibly randomized)
functions \pialice and \pibob, that are used by each
party to operate on its current state $W$ to produce a message $m$ (that is
sent to the other party) and a new state $W'$ for itself. The initial state
of the parties may consist of correlated random variables $(X,Y)$, with
Alice's state being $X$ and Bob's state being $Y$; such a pair is called a set up
for the protocol.
The protocol proceeds by the parties taking turns to apply their respective
functions to their state, and sending the resulting message to the other
party; this message is added to the state of the other party.
$\pi_{\mathrm{Alice}}$ and $\pi_{\mathrm{Bob}}$ also specify when the
protocol terminates and produces output (instead of producing the next
message in the protocol).
A protocol is considered valid only if both parties
terminate in a finite number of rounds (with probability 1).
The {\em view} of a party in an execution of the protocol is a random
variable which is defined as the collection of its states so far in the
protocol execution.
For a valid protocol $\Pi=(\pialice,\pibob)$, we shall denote the final
views of the two parties as $(\Pialice(X,Y),\Pibob(X,Y))$. Also, we shall
denote the outputs as $(\Pialiceout(X,Y),\Pibobout(X,Y))$.

For a protocol
$\Pi$ to be a secure realization of $(U,V)$ given a set up
$(X,Y)$, firstly, the outputs $(\Pialiceout(X,Y),\Pibobout(X,Y))$ must be identically
distributed as $(U,V)$. 
Secondly, if either Alice or Bob is ``curious''
(or ``passively corrupt''), the protocol should give that party no more
information about the other party's output than what their own output
provides. This is formalized using a simulatability requirement. In case
of information theoretic security (as opposed to computational security)
these can be stated in terms of independence of the view, given one's own
output. Formally these three requirements can be stated as follows:%
\footnote{For simplicity, we state the conditions for ``perfect security.''
Our definitions and results generalize to the setting of ``statistical
security,'' where a small statistical error is allowed.}
%
\begin{equation*}
(\Pialiceout(X,Y),\Pibobout(X,Y))  = (U,V)
\end{equation*}
\begin{equation*}
\Pialice(X,Y) \leftrightarrow \Pialiceout(X,Y) \leftrightarrow \Pibobout(X,Y) 
\end{equation*}
\begin{equation*}
\Pialiceout(X,Y) \leftrightarrow \Pibobout(X,Y) \leftrightarrow \Pibob(X,Y) 
\end{equation*}

\subsection{Towards Measuring Cryptographic Content}
In \cite{WolfWu08} three information theoretic quantities were used
to quantify the cryptographic content of a pair of correlated random
variables $X$ and $Y$, which we shall rephrase as below:
{\begin{align*}
H(Y\searrow X|X) &= \min_{Q:H(Q|Y)=I(X;Y|Q)=0} H(Q|X) \\
H(X\searrow Y|Y) &= \min_{Q:H(Q|X)=I(X;Y|Q)=0} H(Q|Y) \\
I(X;Y|X\wedge Y) &= \min_{Q:H(Q|X)=H(Q|Y)=0} I(X;Y|Q)
\end{align*}}
As shown in \cite{WolfWu08}, these quantities are ``monotones'' that  can
only decrease in a protocol, and if the protocol securely realizes a
pair of correlated random variables $(U,V)$ using a set up $(X,Y)$,
then each of these quantities should be at least as large for $(X,Y)$ as for
$(U,V)$.
While these quantities do capture several interesting cryptographic
properties, they paint a partial picture. For instance, two pairs of
correlated random variables $(X,Y)$ and $(X',Y')$ may have vastly different
values for these quantities, even if they are statistically close to each
other, and hence have similar ``cryptographic content.''%

\vspace{0.04cm}
Instead, we shall consider the triplet $\K XYQ$ defined as {\small\[ \K XYQ :=
(I(Q;Y|X), I(Q;X|Y), I(X;Y|Q)),\]} for an arbitrary random variable $Q$.  By
considering all random variables $Q$ we define the region%
\footnote{Here $\le$ stands for coordinate-wise comparison. Note that \KK XY
is equivalent to $\{(\sRRD(R_\RD),R_\RD):R_\RD\in[0, I(X;Y)]\}$. We use this notation
to make the dependence on $X$ and $Y$ explicit.}
{\small\begin{align*}
\KK XY := \{ (x,y,z) \;:\; \exists Q \text{ s.t. } \K XYQ
\le (x,y,z) \}.  
\end{align*}}
This generalizes the three quantities considered in \cite{WolfWu08}, as
(using arguments similar to that used for \corollaryref{GacsKo}) it can be
shown that the region $\KK XY \subseteq \Rplus^3$ intersects the co-ordinate
axes at the points $(H(Y\searrow{X}|X),0,0)$, $(0,H(X\searrow{Y}|Y),0)$, and
$(0,0,I(X;Y|X\wedge Y)$.  In the following sections we point out that \Kfunc
also satisfies a monotonicity property: the region can only expand in a
protocol, and if the protocol securely realizes a pair of correlated random
variables $(U,V)$ using a set up $(X,Y)$, then \KK XY should be smaller than
\KK UV.  As we shall see, since the region \KK XY has a non-trivial shape
(see for instance, \exampleref{zsource}), \Kfunc can yield much better bounds on the
rate than just considering the axis intercepts; in particular \Kfunc can
differentiate between pairs of correlated random variables that have the
same axis intercepts.  Further \KK XY is continuous as a function of
$(X,Y)$, and as such one can derive bounds on rate that are applicable to
statistical security as well as perfect security.

\subsection{Monotone Regions for 2-Party Secure Protocols}
\label{sec:monotone}
Given a pair of random variables $(X,Y)$ denoting the {\em views} of the two
parties in a 2-party protocol we are interested in capturing the
``cryptographic content'' of this pair. We shall do so by defining a
region in multi-dimensional real space, that  intuitively, consists of witnesses
of ``weakness'' in the cryptographic nature of the random variables $(X,Y)$;
thus smaller this region, the more cryptographically useful the variables
are.  The region has a monotonicity property: a secure protocol that
involves only communication (over noiseless links) and local computations
(i.e., without using trusted third parties) can only {\em enlarge} the region.

Our definition of a monotone region from \cite{MajiPrPrRo10} given below, strictly generalizes that suggested by
\cite{WolfWu08}. The monotone in \cite{WolfWu08}, which is
a single real number $m$, can be interpreted as a one-dimensional region
$[m,\infty)$ to fit our definition. (Note that a decrease in the value of
$m$ corresponds to the region $[m,\infty)$ enlarging.)
\begin{defn}
\label{def:monotone}
We will call a function \Mfunc that maps a pair of random variables $X$ and
$Y$, to an upward closed subset%
\footnote{A subset \Mfunc of $\Real^d$ is called upward closed if $\pt \in
\Mfunc$ and $\pt' \ge \pt$ (i.e., each co-ordinate of $\pt'$ is no less than that
of \pt) implies that $a'\in \Mfunc$.}
of $\Rplus^d$ (points in
the $d$-dimensional real space with non-negative co-ordinates)  a {\em monotone region}
if it satisfies the following properties:
\begin{enumerate}
\item ({\em Local computation cannot shrink it.})
For all random variables $(X,Y,Z)$ with $X \leftrightarrow Y
\leftrightarrow Z$, we have
$ \M{XY}{Z} \supseteq \M{Y}{Z}$ and $\M{X}{YZ} \supseteq \M{X}{Y}$.

\item ({\em Communication cannot shrink it.})
For all random variables $(X,Y)$ and functions $f$ (over the support of
$X$ or $Y$), we have
$ \M{X}{Yf(X)} \supseteq \M{X}{Y}$ and $\M{Xf(Y)}{Y} \supseteq \M{X}{Y}$.

\item  ({\em Securely derived outputs do not have smaller regions.})
For all random variables $(X,U,V,Y)$ with $X \leftrightarrow U \leftrightarrow V$ and
$U\leftrightarrow V \leftrightarrow Y$, we have
$ \M{U}{V} \supseteq \M{XU}{YV}$.

\item ({\em Cryptographic content in independent pairs add up.})
For independent pairs of random variables $(X_0,Y_0)$ and $(X_1,Y_1)$, we
have
$ \M{X_0X_1}{Y_0Y_1} = \M{X_0}{Y_0} + \M{X_1}{Y_1}, $
where the $+$ sign denotes {\em Minkowski sum}. That is,
$\M{X_0X_1}{Y_0Y_1} = \{ \pt_0+\pt_1 \;|\; \pt_0 \in \M{X_0}{Y_0} 
 \text{ and } \pt_1\in\M{X_1}{Y_1} \}$.
(Here addition denotes coordinate-wise addition.) 
\end{enumerate}
\end{defn}
Note that since \M{X_0}{Y_0} and \M{X_1}{Y_1} have non-negative co-ordinates
and are upward closed, $\M{X_0}{Y_0} + \M{X_1}{Y_1}$ is smaller than both of
them.  This is consistent with the intuition that more cryptographic content
(as would be the case with having more independent copies of the random
variables) corresponds to a smaller region.

\subsection{\Kfunc as a Monotone Region.}
\label{sec:KKmonotone}
In \cite{MajiPrPrRo10} we prove the theorem below, and obtain
the following corollary.
\begin{thm}
\Kfunc is a monotone region as defined in \definitionref{monotone}.
\end{thm}
\begin{corol} \label{cor:secure-realization-rate}
If $n_1$ independent copies of a pair of correlated random variables $(U,V)$ can be securely realized
from $n_2$ independent copies of a pair of correlated random variables $(X,Y)$, then
$n_1 \KK XY \subseteq n_2 \KK UV$. (Here multiplication by an integer $n$
refers to $n$-times repeated Minkowski sum.)
\end{corol}

Intuitively, \KK XY captures the cryptographic content of the correlated
random variables $(X,Y)$: the farther it is from the origin, the more
cryptographic content it has. In particular, if \KK XY contains the origin,
then $(X,Y)$ is cryptographically ``trivial,'' in the sense that $(X,Y)$ can
be securely realized with no set ups.
%
%
This triviality property can be inferred from the three
quantities considered by \cite{WolfWu08} as well, since those
quantities correspond to the axis intercepts of our monotone region.
However, what makes the monotone region more interesting is when the pair of
correlated random variables is non-trivial, as illustrated in the following
example.

\begin{eg} \label{eg:crypto-example} Consider the question of securely
realizing $n_1$ independent pairs of random variables distributed according
to $(U,V)$ in \exampleref{zsource} from $n_2$ independent pairs of $(X,Y)$
in \exampleref{connected}.  While the monotones in \cite{WolfWu08} will give
a lowerbound of 0.5182 on $n_2/n_1$, we show that $n_2/n_1 \ge 1.8161$. (For
this we use the intersection of \KK UV with the plane $z=0$
(\figureref{zsource}) and one point in the region \KK XY (marked in
\figureref{connected}), and apply \corollaryref{secure-realization-rate}.)
\end{eg} 

Hence, the axis intercepts of this monotone region (one of which is the common
information of G\'{a}cs and K\"{o}rner) do not by themselves capture subtle
characteristics of correlation that are reflected in {\em the shape of the
monotone region}. As discussed in \cite{MajiPrPrRo10}, \KK XY is a convex
region, and for a fixed set of axis intercepts, the cryptographic quality of
a pair of random variables is reflected in how little it bulges towards the
origin. We leave as an open question whether our bound is indeed tight.

\end{document}